\documentstyle[prl,aps,psfig]{revtex}
\draft
\newcommand{\be}{\begin{equation}}
\newcommand{\ee}{\end{equation}}

\newcommand{\bary}{\begin{eqnarray}}
\newcommand{\eary}{\end{eqnarray}}
\def \gnuph {{g_{\nu\phi}}}
\def \gnupp {{g^2_{\nu\phi}}}
\begin{document}
\title{Emission of electron density waves by neutrinos in a dense medium} 
\author{
Subhendra Mohanty$^{*}$, Sarira Sahu$^{\dag}$ and Srubabati Goswami$^{a}$}
\address{ Physical Research Laboratory,
Navrangpura, Ahmedabad - 380 009, India\\
and\\
$^a$Calcutta University, Calcutta, India}
\maketitle

\begin{abstract}
We study the production of electron density waves by neutrinos propagating 
through a plasma. We treat this process in field theoretically as a Cerenkov
emission of phonons (quanta of electron density waves) by neutrinos. We 
compute the energy transfered to the plasma by the neutrinos and apply it to
the shock wave revival problem of supernovae.
\end{abstract}

\section{Introduction}
High energy particles propagating in a dense plasma can loose their kinetic
energy by perturbing the particle distribution of the plasma. The
fluctuation in the number densities of protons and neutrons is small as they
are heavy but for electrons the fluctuations are significant. Neutrinos 
propagating through a plasma will scatter electrons by weak interactions.
The primary energy transfer process is inelastic scattering of neutrinos
by nucleons ($\nu_e n\rightarrow e^-p$) which was first studied by
Bethe and Wilson\cite{bw}. The Bethe-Wilson process was introduced as a 
mechanism for transferring energy to the shock wave of supernovae which
would enable it to overcome the gravity of the supernova core\cite{woosley}. 
The energy delivered by the Bethe-Wilson process 
is however not enough to revive
the stalled shock wave\cite{woosley}. 

  Bingham et al.,\cite{shukla2} have introduced a semiclassical 
mechanism analogous to
the electromagnetic pondermotive force where by a inhomogeneous flux of 
neutrinos through plasma transfers energy collectively to the  medium.
The energy transfer by the neutrino pondermotive force is sufficient for
the shock wave explosion of supernovae. A field theoretic derivations of
this concept is not yet been achieved\cite{melrose}
   
      In this paper we study another collective process whereby neutrinos can
transfer energy to plasma through which they propagate. We study the 
generation of electron density waves by the passage of neutrinos. In field
theoretic terms we study the emission of 'phonons' (the quanta of electron
density waves) by neutrinos by the weak interaction process. Such a single
vertex emission of a quanta which is possible only in a medium is called
a Cerenkov process\cite{grimus,mohanty,pal1,sarira,raffelt}
This is kinematically allowed when the emitted
'particles' have a spacelike dispersion relation ($k^2 > \omega^2$) in
the medium. 

 We show that for a range of frequencies the refractive index of the electron density waves $n(\omega) > 1$ and therefore the Cerenkov emission of these
waves is kinematically allowed. We compute the rate of emission of
phonons and the energy transfer by phonon emission from neutrinos.
We find that the energy 
transfer by the Cerenkov emission of phonons is less than the energy 
transfered by the Bethe-Wilson inelastic scattering process.

The paper is organised in the following manner. In section II we derive 
the dispersion relation of the electron density waves in a plasma. The
Cerenkov emission of phonon is considered in section III. We calculate the
amount of energy deposited by the Cerenkov emission of phonon, in the
stalled supernova medium and compare this with the BW mechanism in
section IV. Our results are briefly discussed in conclusions.

\section{Phonon  dispersion relation}

The Lagrangian for the elastic electron electron scattering is given by
\be
{\cal{L}}_{int} = \frac{e^2}{K^2} (\bar{u}_e \gamma_{\mu} u_e)
(\bar{u}_e \gamma^{\mu} u_e )
\ee
Now considering free electron scatter from the background electrons
(electron plasma) then,
we have to average the background electrons which will give
\be
{\cal{L}}_{int} = \frac{e^2}{K^2} \bar{u}_e \gamma_{\mu} u_e
\langle \bar{u}_e \gamma^{\mu} u_e \rangle
\ee
where, 
$\langle \bar{u}_e \gamma^{\mu} u_e \rangle = n^{\mu}$ denotes
the averaging over the background electrons and $n^{\mu}=(n_e, {\bf j})$. 
The quantity $n_e$ is the mean electron density in the medium and
$\bf j$ is the current density. Fluctuation in $n^{\mu}$ is 
given by
$n^{\mu} + \delta n^{\mu}$, where $\delta n^{\mu}$ is 
the fluctuation over the mean $n_{\mu}$.
We are interested in the part of the Lagrangian, which corresponds to
the coupling of the $\delta n^{\mu}$ field with the electron. 
From the above equation, we obtain for the fluctuation part
\be
{\delta\cal{L}}_{int} = \frac{e^2 n_e}{K^2 m_e} \bar{u}_e \gamma_\mu u_e
\left (\frac {\delta{n^{\mu}} m_e} {n_e}\right ) 
\label{coupling}
\ee
where ${\delta{n^\mu}m_e}/{n_e}$ defines the phonon field $\phi^{\mu}$.
From eq.(\ref{coupling}) the effective coupling of the field $\phi$
with the electrons, is given by
\vspace*{2.3in}
\begin{figure}
\includegraphics{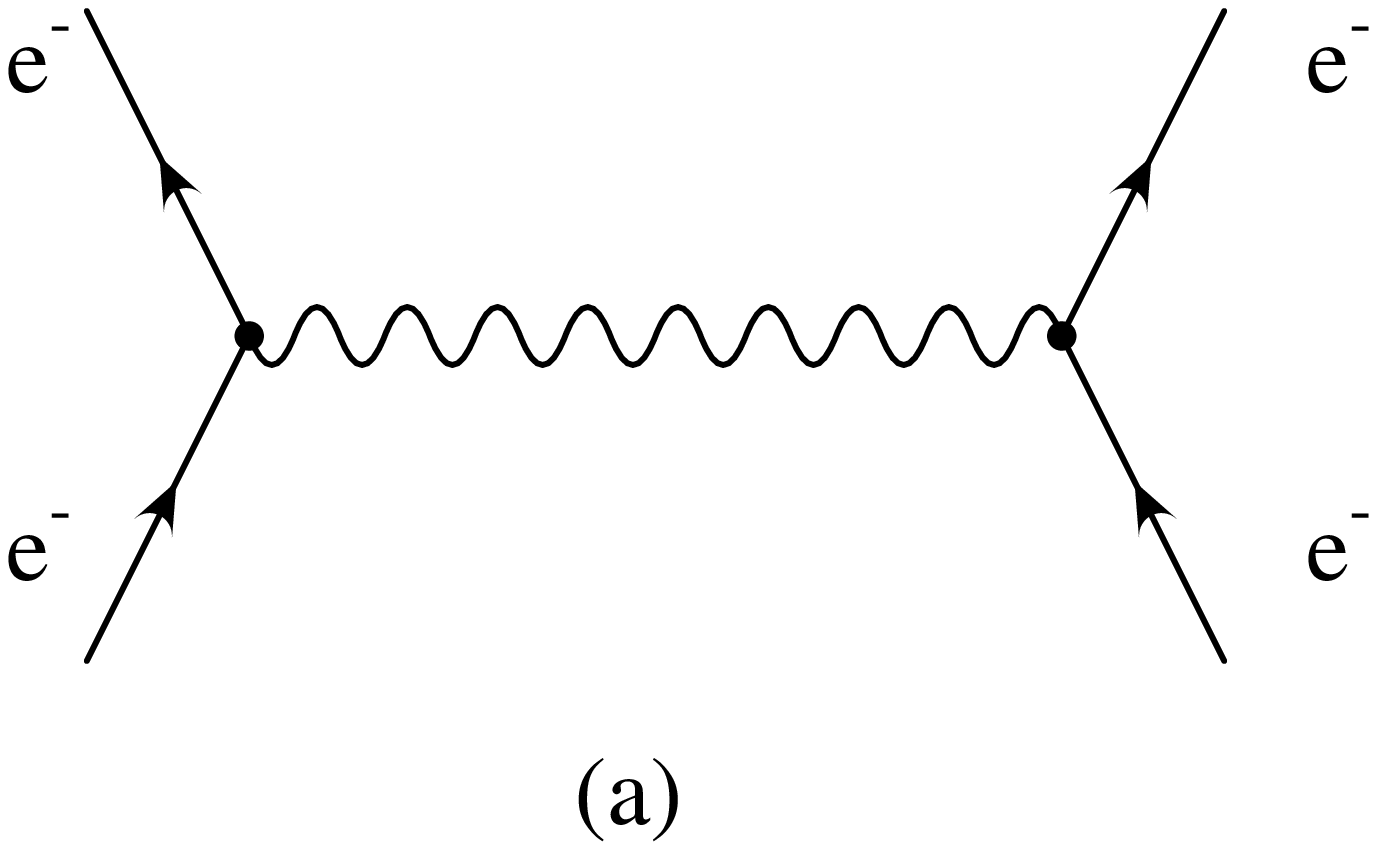}
\includegraphics{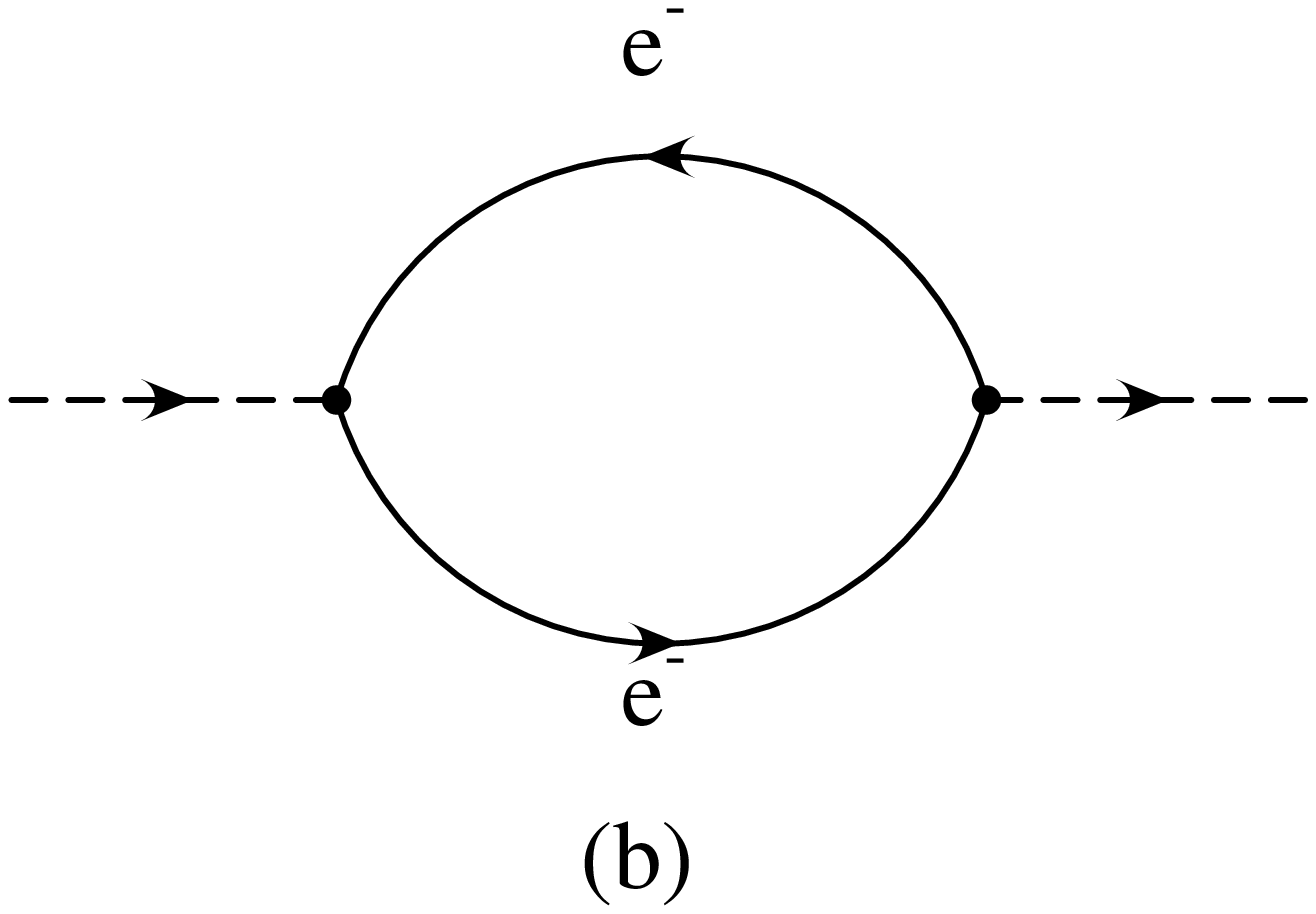}
\end{figure}
\vskip -3cm
\begin{center}
{\it Figure 1: (a) 
Feynman diagram for 
$e^- + e^-\leftrightarrow e^- + e^-$  (b) Phonon polarisation in the medium}
\end{center}
\be
g = \frac{e^2 n_e}{K^2 m_e}=\frac{4\pi\alpha n_e}{m_e K^2}.
\label{g}
\ee
We are now in a position to evaluate fig.1 
It gives,
\be
i\Pi_{00}(K) = - g^2 {\int {\frac {d^4 p} { {(2\pi)}^4 } 
Tr\left [ \gamma_{0} {S_F}(p) \gamma_{0} S_{F}(p+K) \right ]}}
\label{pi1}
\ee
where, $S_{F} (p)$ is the electron propagator at finite
temperature and density. Using the real time formalism of the
finite-temperature field theory\cite{dolan,weldon}, the electron 
propagator in the medium is given by 
\be
S_{F}(p) = (\not{p} + m)\left [\frac{1}{p^2 - m^2} + 2\pi i
\delta(p^2 - m^2) {f_F}(p.u)\right ]
\label{sf}
\ee
where $u_{\mu}$ is the four-velocity of the center of mass of
the medium and $f_F$ denotes the Fermi distribution function
\be
f_{F}(x) = \frac {\theta(x)}{e^{\beta(x-\mu)}+1} + 
\frac{\theta(-x)}{e^{-\beta(x-\mu)}+1}. 
\ee
Here, $\theta$ is the unit step function, $\beta$ is the inverse
temperature and $\mu$ denotes the chemical potential of the
electron. 
After carrying out the trace and computing the integral over
$d^3 p$ eq.(\ref{pi1}) becomes,
\bary
\Pi_{00}(K) & = & -\frac {g^2} {2 \pi^2 k}
\int  dp_{0} f_{F} (p_0) 
\left [
\left ( p_{0}^2 + p_{0}\omega  + {K^2\over 4} \right )
ln \left ( \frac{K^2 + 2p_{0}\omega +2\sigma k} 
{K^2 + 2p_{0}\omega - 2\sigma k} \right )\right .
\nonumber \\
& & \left .+ 
\left (p_{0}^2 - p_{0}\omega  + {K^2\over 4}\right )
ln \left (
\frac {K^2 - 2p_{0}\omega +2\sigma k} {K^2 - 2p_{0}\omega -2\sigma k} 
\right ) - 2\sigma k \right ]
\label{pito}
\eary
where, $p_{\mu}= (p_{0},{\bf p})$, $k_{\mu} = K = (\omega, {\bf
k})$, $\mid{\bf{k}}\mid = k = n\omega$, $n$ denoting the
refractive index for the phonon and 
$\sigma = (p_{0}^2 - m_e^2)^{1/2}$.
In the limit $p_0 >> m_e$ and $\omega$, the lograthmic terms in eq.(\ref{pito})
will be simplified to
\be
ln\left (\frac{K^2+2 p_0\omega +2\sigma k}
{K^2+2 p_0\omega -2\sigma k} \right )
\simeq ln\left (\frac{\omega + k}{\omega-k}\right ) -\frac{k}{p_0} + 
\frac{\omega k}{2 p_0^2} + ...
\ee
and
\be
ln\left (\frac{K^2-2 p_0\omega +2\sigma k}
{K^2-2 p_0\omega -2\sigma k} \right )
\simeq -ln\left (\frac{\omega + k}{\omega - k}\right ) -\frac{k}{p_0} - 
\frac{\omega k}{2 p_0^2} + ...
\ee
Keeping only the leading order terms in $p_0$ in eq.(\ref{pito}) we obtain
\be
\Pi_{00}=-\frac{2 g^2}{\pi^2}\int dp_0 p_0 f_F(p_0) 
\left (\frac{\omega}{2 k} 
ln\left (\frac{\omega +k}{\omega - k}\right ) -1\right )
\ee
The Fermi momentum is given by
$p_{Fe}=3\pi^2 n_e$. For matter of density $\rho\simeq 10^8~ gm/cm^3$
the electron Fermi momentum is approximately $1.6 MeV$. So the electron
chemical potential $\mu\simeq p_{Fe}=1.6~ MeV$. 
Thus we consider
$p_0 >> \mu$ here. 
Also in the stalled shock wave medium there is no positron. 
With the above approximation we obtain, 
\be
Re{\Pi_{00}} = -\frac{g^2 T^2}{6} 
\left (\frac{\omega}{2 k} 
ln\left |\frac{\omega + k}{\omega-k}\right | - 1\right ),
\ee  
where the plasma has the temperature $T$ of order MeV.
The dispersion relation satisfied by the phonon is
\be
w^2-k^2-Re\Pi_{00}=0
\label{ds}
\ee
Putting the value of $g$ from eq.(\ref{g}) in eq.(\ref{ds}) we obtain
the following dispersion relation
\be
(n^2-1)^3=\left (\frac{4\pi\alpha n_e}{m_e}\right )^2 
\frac{T^2}{6 \omega^6}
\left (\frac{1}{2 n}\ln\left |\frac{1+n}{1-n}\right |-1\right ).
\label{disp}
\ee
 Figure 3, shows that for the refractive index $n$ in the range
$1 < n < 1.2$ the phonon frequency $\omega$ is positive. 
On the other hand for $n > 1.2$ the phonon frequency is no longer positive
and phonon propagation is Landau damped. We are interested only in the
range of refractive index $n > 1$ for which phonon frequency is positive,
because this is the range in which Cerenkov emission of phonon 
is kinematically allowed.
\vspace*{3.4in}
\begin{figure}
\includegraphics{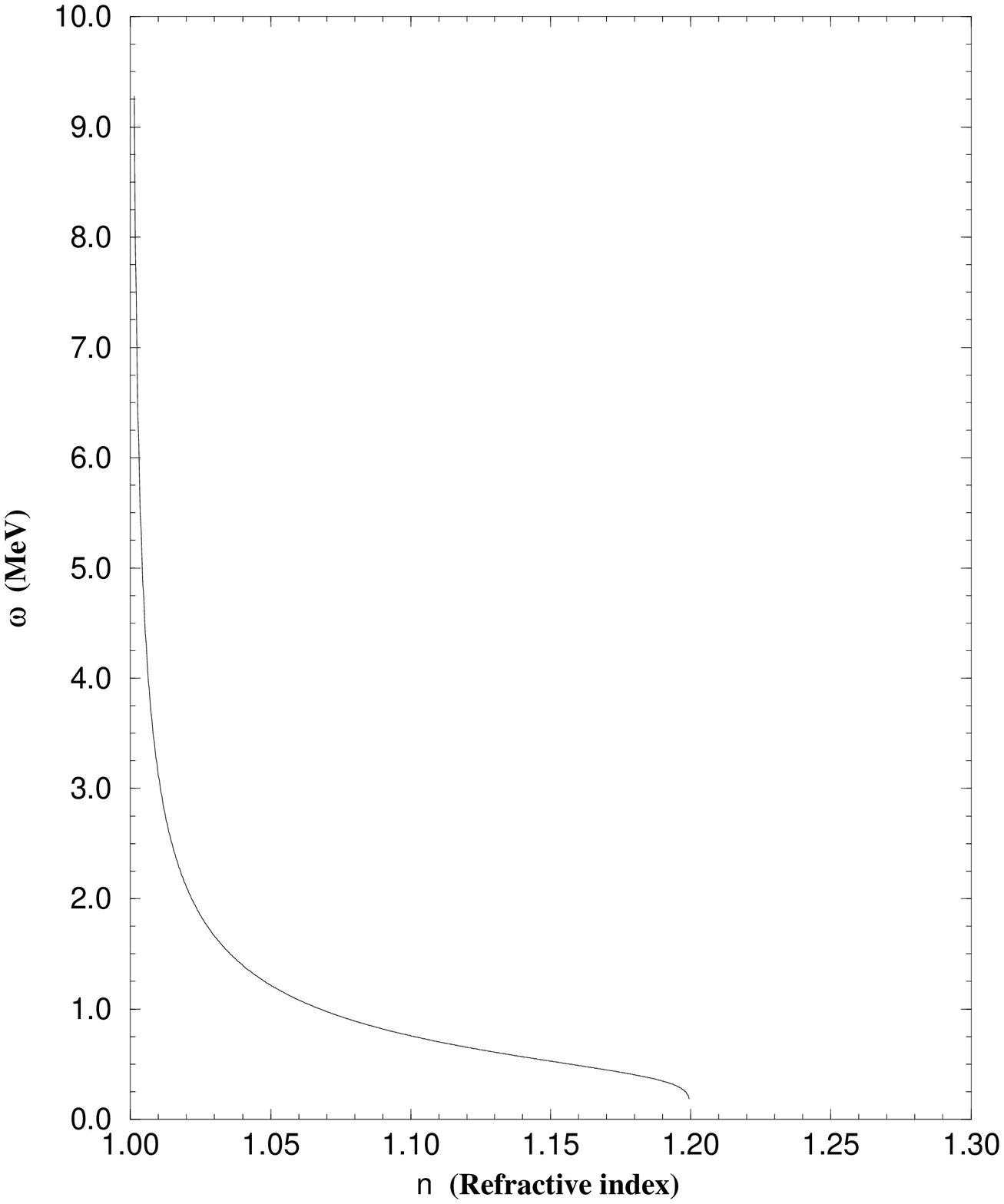}
\label{fg:feynr}
\end{figure}
\vspace*{0.3in}
\begin{center}
{\it Figure 2: Dispersion relation for the phonon}\\ 
\end{center}

\section{Cerenkov Radiation of Phonon}

For neutrino electron scattering the Lagrangian is given by
\be
{\cal L} = {G_F\over \sqrt 2}{\bar u}_{\nu_e}\gamma_{\mu} 
(1-\gamma_5) u_{\nu_e}
{\bar u}_{e}\gamma_{\mu} (1-\gamma_5) u_{e},
\ee
where $G_F=1.166\times 10^{-5}/GeV^2$ is the Fermi coupling constant.
For the background electrons we average
$\langle \bar{u}_e \gamma_{\mu} u_e \rangle$ 
(assuming the electrons to be unpolarised).
Then perturbing the electron density as done in eq.(\ref{coupling}), 
we obtain the effective neutrino-phonon interaction Lagrangian as
\be
{\cal L}_{\nu\phi}={i G_F n_e\over {\sqrt{2} m_e}}
{\bar u}_{\nu_e}\gamma_{0} (1-\gamma_5)u_{\nu_e}\phi,
\label{lag}
\ee
where $\phi$ is the field defined in the previous section.
Eq.(\ref{lag}) gives the effective neutrino-phonon coupling 
as
\be
{\gnuph}={G_F n_e\over {{\sqrt 2} m_e}}
\ee
Let us consider the $\phi$ emission process from neutrino in the medium
\be
\nu_e(p_1)\rightarrow\nu_e(p_2) + \phi(K), 
\label{proc}
\ee
where $p_1=(E_1, {\bf p_1})$, 
$p_2=(E_2, {\bf p_2})$ and $K$ are the four-momenta of 
incoming neutrino, outgoing neutrino and outgoing phonon respectively.
The matrix element for the  process in eq.(\ref{proc}) is
\be
i{\cal M} = 
i\gnuph \epsilon_{\mu} {\bar u}_{\nu_e}(p_2)\gamma_{\mu}(1-\gamma_5)
u_{\nu_e} (p_1),
\ee
and after averaging over the initial neutrino spin this gives 
\be
|{\cal M}|^2 = 4{\gnupp}\sum\epsilon_{\mu}\epsilon^*_{\nu}
\left ( p_{2\mu}p_{1\nu} +  p_{1\mu}p_{2\nu} -  
p_{1}p_{2} g_{\mu\nu}\right ). 
\label{mm}
\ee
Here $\epsilon_{\mu}$ is the polarisation vector for the field $\phi$.
As we consider only the longitudinal mode, the polarisation sum will also
be on the longitudinal mode only.
The polarisation sum for longitudinal mode in the medium is given by
\be
\sum \epsilon^L_{\mu}(k)
\epsilon^{L*}_{\nu} (k)=
{1\over {n^2\omega }} (u_{\mu}k_{\nu}  + k_{\mu} u_{\nu})
-{1\over {n^2 {\omega^2}}} k_{\mu}k_{\nu},
\label{pol}
\ee
with $u_{\mu}=(1, 0)$ the center of mass velocity of the medium.
Using eq.(\ref{pol}) in eq.(\ref{mm}), $|{\cal M}|^2$
is given by
\be
|{\cal M}|^2 = {4{\gnupp}\over n^2}
E_1\left [2 E_1 + 2 E_1 n^2 - \omega + n^2\omega 
+ n\cos\theta \left (1-4 E_1 - n^2\omega\right ) \right ]
\ee
The total energy emitted from a single process is
\be
S={1\over {2 E_1}}\int {d^3p_2\over {2 E_2 (2\pi)^3}}
{d^3k\over {2 \omega (2\pi)^3}} 
\omega (2\pi)^4 \delta^4 (p_1 - p_2-K) |{\cal M}|^2.
\label{ss}
\ee
We use the identity
\be
\int {d^3p_2\over {2 E_2}} =
\int d^4p_2 \Theta (E_2) \delta(p_2^2 - m_{\nu}^2),
\label{dp4}
\ee
where $\Theta (E_2)$ is the step function and $m_{\nu}$ is the
neutrino mass. Putting eq.(\ref{dp4}) in 
eq.(\ref{ss}) and integrating over $d^4p_2$ we obtain the rate of energy 
radiated per neutrino is
\be
{\dot S}={1\over {16\pi^2 E_1}}\int {d^3k\over {2{\bf |p_1|}k}}
\delta\Big ( 
{{(2E_1 \omega-\omega^2 +k^2)}\over {2 {\bf|p_1|}k}} - cos\theta\Big )
|{\cal M}|^2.
\label{si}
\ee
The angle $\theta$ between the incoming neutrino and 
the emitted phonon is obtained from the delta function in eq.(\ref{si}),
\be
\cos\theta = {(2 E_1 \omega - \omega^2 + k^2)\over {2{\bf |p_1|}k}}
={1\over {nv}}\Big (1 + {(n^2-1) \omega\over {2 E_1}}\Big ),
\label{cos}
\ee
where $v =|{\bf p_1}|/ E_1$ is the neutrino velocity ($\simeq$ 1). 
Neutrino mass term being small we neglect it. Since
$-1 \le cos\theta \le 1$; which implies
\be
-{2 E_1\over (n-1)}\le \omega \le {2 E_1\over (n+1)}.
\ee
But definitely $-{2 E_1/ (n-1)}$ can not be the lower limit for the 
above process as for $n >1$ this is a negative quantity and 
$\omega$ can not be negative. The lower limit for $\omega$ will be calculated
from the dispersion relation obtained for phonon in the previous section.
Figure 3 shows that, the phonon frequency is minimum for higher value of
the refractive index and for $n\simeq 1.2$, $\omega=\omega_1\simeq0.19~MeV$.
and the upper limit depends on the value of the incoming neutrino energy.
Putting the value of $\cos\theta$ from the eq.(\ref{cos}) in $|{\cal M}|^2$
and simplifying eq.(\ref{si})we get
\be
{\dot S}={{\gnupp}\over {16\pi E_1^2}} \int_{\omega_1}^{\omega_2}
(n^2-1) \omega d\omega 
\left [4 E_1^2 - 4 E_1 \omega + \omega^2 (n^2 - 1)\right ].
\label{sdot}
\ee
This shows that only for $n > 1$ the phonon emission is possible, which is
the usual Cerenkov condition.
Substituting  the value of $(n^2-1)$ from the phonon dispersion relation
eq.(\ref{disp}) in  eq.(\ref{sdot}) one obtains,
\be
{\dot S} =\frac{G_F^2 n_e^2}{16\pi m_e^2 E_1^2}
\left (\frac{4\pi\alpha n_e}{m_e}\right )^{2/3}
\left (\frac{T^2}{6}\right )^{1/3}
\int_{\omega_1}^{\omega_2} \frac{d\omega}{\omega} 
f(n) \left [ 4 E_1^2 - 4 E_1\omega +
\left (\frac{4\pi\alpha n_e}{m_e}\right )^{2/3}
\left (\frac{T^2}{6}\right )^{1/3}f(n) 
\right ],
\label{master}
\ee
where we have defined 
\be
f(n) =
\left [ \frac{1}{2n}ln\left |{{1+n}\over {1-n}}\right |-1\right ]^{1/3}.
\ee
The integral in eq.(\ref{master}) can not be evaluated analytically. So
we have evaluated this numerically. 
For $\rho=10^8 g/cm^3$, $T=2 MeV$ and $E_1=10 MeV$ we obtain
${\dot S}=3.23\times 10^{-24}~MeV^2$.

\section{Supernova shock revival}

Type-II supernova are consequence of the collapse of the iron core of 
massive stars of $8 M_{\odot}\le M \le 25 M_{\odot}$ and lead to the formation
of a neutron star or black hole. Observations of neutrino events from 
supernova SN1987A explosion, by Kamiokande II and IMB detectors have 
confirmed the
fundamental aspects of the theoretical understanding of type-II supernovae.
However the mechanism of causing the supernova explosion is yet to be
understood satisfactorily.

  The neutrino streaming up from deeper region of the supernova are 
supposed to deposit a small fraction of their energy in the matter between
the protoneutron star and the stalled shock, which is about 100-200 Km 
away from the core. 
Recent numerical calculations in more than one dimension shows that material
behind the stalled shock wave of the supernova can be heated efficiently
by the neutrinos coming from the neutrinosphere and
eventually expel the outer mantle
causing the supernova explosion\cite{arnet,burrows,fuller,akhmedov}.
Matter-enhanced neutrino oscillation (MSW) effect in supernova is considered
for the shock revival. The fact that, the region between the neutrino sphere
and the stalled wave density is such that flavour transformation of
$\nu_{\mu}$ or $\nu_{\tau}$ to $\nu_e$ is resonant for massive neutrinos.
Since the average energy of $\nu_{\mu}$'s and $\nu_{\tau}$'s at the neutrino
sphere is about 20 MeV whereas that of $\nu_e$'s is about 10 MeV. The
oscillation of $\nu_{\mu}(\nu_{\tau}$)to $\nu_e$ would have twice as high 
energy as the originally emitted ones, and this extra energy would be available
for heating the matter behind the shock. Also it has been proposed that
spin-flavour precession of neutrinos may play an important role in the 
explosion of the stalled wave. In particular, it can be resonantly enhanced 
in the region between the neutrino sphere and the stalled matter.
But it is still controversial, whether the neutrino 
energy is sufficient for strong enough heating to revive the stalled 
shock wave.

 Bethe and Wilson (BW) in 1985 showed that\cite{bw}, 
neutrinos from the hot inner core
of the supernova are captured by the matter behind the shock through the
process $\nu_e +n\rightarrow p +e^-$ and ${\bar\nu_e} + p\rightarrow n + e^-$
and deliver their energy. It was argued by BW that $0.1\%$ of the total energy
of the neutrino and anti-neutrino capture process is sufficient to reheat
the stalled shock wave and cause supernova explosion.

  In a recent paper by one of us\cite{sarira}, 
has shown that, neutrinos propagating in
the stalled medium, will emit longitudinal photons (plasmons) by Cerenkov
process and deposit some energy in the stalled matter, while transverse
photons emission is not possible because the transverse mode is Landau
damped. Comparison of the plasmon emission process with BW process shows that
the former one is a very weak process to account for the shock heating.

Here we are interested to calculate the amount of energy emitted due to
Cerenkov emission of phonon, which is subsequently absorbed by the matter
from the neutrinos which are propagating through this stalled medium and 
compare with the BW process.

 The total energy deposited by Cerenkov emission of phonons by neutrinos per
unit time within the stalled matter of thickness $d$ is given by
\be
{\dot E}_p = {\dot S}~ d ~(Neutrino ~Flux)
\ee
where ${\dot S}$ is given in eq.(\ref{master}) and neutrino flux is 
$10^{52}~ erg/sec/E_1$. For stalled shock 
density $\rho=10^8~gm/cm^3$\cite{woosley},
temperature 2 MeV and neutrino energy $E_1=10~ MeV$
we obtain ${\dot S}=3.23\times 10^{-24}~MeV^2$. This gives
\be
{\dot E}_p = 1.64\times 10^{38}~ d_{cm}~ erg/sec.
\ee

In BW mechanism
rate of energy absorbed
by a gram of matter at a distance R is\cite{bw}
\be
{\dot E_{BW}}=3\times 10^{18} L_{\nu 52}
\Big ( {T^2_{\nu}\over R^2_7}\Big )~
{\tilde Y_N}~~ erg/g/sec,
\ee
where $L_{\nu 52}$ is the neutrino luminosity
in units of $10^{52}$ erg/sec,
$R_7$ is the distance from the center in units of $10^7$ cm, $T_{\nu}= 5$
MeV is the temperature of the neutrino sphere and 
$\tilde Y_N \simeq 1$
is the total mean fraction of the nucleon. Here we neglect the contribution
due to electron and positron capture as they are correction to this
contribution. The total energy absorbed by the stalled shock wave,
which has a mass $4\pi\rho R^2 d$ (having thickness $d$ and
density $\rho$)  is
\be
{\dot E_{BW}} = 3\times 10^{18}~ L_{\nu 52} ~
\Big ({T^2_{\nu}\over R^2_7}\Big )
~~{\tilde Y_N}~~ erg/g/sec \times 4\pi\rho R^2 d.
\label{bw}
\ee
For the neutrino luminosity $L_{\nu_e} = 10^{53}~~erg/sec$
and the stalled
shock wave density $\rho\simeq 10^8~ g/cm^3$ and $R =200$ Km,
the energy absorbed by the shock wave (assuming 100\% absorption) is
\be
{\dot E_{BW}}= 9.4\times 10^{43}~ d_{cm}~~ erg/sec
\ee
where $d_{cm}$ is units of cm.
Comparing phonon emission process with BW gives
\be
\frac{{\dot E}_p}{{\dot E}_{BW}}\simeq 10^{-6}.
\ee
So this shows that phonon emission  by electron density disturbance
due to neutrino does not deposit as much energy as the BW process.

\section{conclusions}

 We have studied the generation of electron density waves by neutrinos
propagating in the electron plasma of stalled shock wave of type-II 
supernovae. In the field theoretical language, we treat this density 
wave as the quanta of phonon. Calculation of dispersion relation for
the phonons show that, for a narrow range of the refractive index 
$1 < n \le 1.2$ we can have Cerenkov emission of phonon quanta in the
stalled shock wave medium of the supernovae. These phonons are subsequently
absorbed by the medium, thus depositing some of the neutrino energy.
We found that, this process is much weaker than the BW process.
The phonon contribution is small because of two reasons. Firstly
the emission rate is proportional to the square of the Fermi coupling 
constant, which is a small number  and 
secondly the integral in eq.(\ref{master}) is small, because 
most of the phonons are emitted in the low energy range (small $\omega$)
as shown in Figure2.

\end{document}